\documentstyle[aps,preprint]{revtex}
\begin{document}
\draft
\tightenlines

\title{Semiclassical evaluation of average nuclear
\\
one and two body matrix elements}

\author{X. Vi\~nas$^1$, P. Schuck$^2$, M. Farine$^3$
and M. Centelles$^1$}

\address{
$^1${\it Departament d'Estructura i Constituents de la Mat\`eria,
Facultat de F\'{\i}sica,
\\
Universitat de Barcelona,
Diagonal {\sl 647}, {\sl 08028} Barcelona, Spain}
\\
$^2${\it Institut de Physique Nucl\'eaire, IN{\sl 2}P{\sl 3}--CNRS,
\\
Universit\'e Paris--Sud, {\sl 91406} Orsay-C\'edex, France}
\\
$^3${\it Consulat G\'en\'eral de France \`a Canton, 
{\sl 339} Huan Shi Dong Lu,} 
\\
{\sl 510098} Guangzhou, Canton, China}


\maketitle

\begin{abstract}

Thomas-Fermi theory is developed to evaluate nuclear matrix elements
averaged on the energy shell, on the basis of independent particle
Hamiltonians. One- and two-body matrix elements are compared with the
quantal results and it is demonstrated that the semiclassical matrix
elements, as function of energy, well pass through the average of the
scattered quantum values. For the one-body matrix elements it is shown
how the Thomas-Fermi approach can be projected on good parity and also
on good angular momentum. For the two-body case the pairing matrix
elements are considered explicitly.
\end{abstract}

\vspace*{1cm}




\section{Introduction}

The solution of the nuclear many-body problem presents a formidable
challenge. Not only bare and effective nucleon-nucleon forces are not
completely known, but still for those given as granted one has to
solve the many-body problem of a highly quantal, strongly interacting,
self-bound, and therefore inhomogeneous Fermi system. Over the years
semiclassical techniques have helped to solve this problem, for
instance in regard to the latter aspect. In practice it is mainly the
Thomas-Fermi (TF) method and its extensions for the description of
nuclear ground-state properties which has been considered (see
\cite{R1} and references therein). The nuclear density and kinetic
energy density are the main ingredients of this approach.

The semiclassical approximation 
often gives a direct physical insight, yielding the shell average of
the quantities under consideration and providing their main trend
(which, in certain cases, may be obscured by strong shell
fluctuations).
A known example is e.g. the nuclear binding energy which coincides
with the liquid drop part in the semiclassical approach. Another
quantity of longstanding interest is the average single-particle level
density. It is well known that the TF approximation to the level
density (including $\hbar$ corrections) coincides analytically with
the Strutinsky averaged quantal level density for the harmonic
oscillator (HO) potential \cite{R01}. First performing the quantal
calculation and then the average is more cumbersome than calculating
the shell average directly via the TF method. The technical advantage
of the latter becomes significant in the deformed case \cite{R01} or,
for  instance, when one wants to go beyond the independent particle
description to include correlations \cite{R02}. The TF approach is
also very helpful for the calculation of surface and curvature
energies. Actually, the latter quantity can only be correctly
extracted in a semiclassical procedure \cite{R03}.

In general, however, we believe that the true virtue of the TF method
shows up not only in calculating average properties in the independent
particle approximation, where it can replace the results obtained
through the more cumbersome Strutinsky method \cite{R04}, but rather
in many-body applications going beyond the mean field or independent
particle picture where a straightforward quantum solution for finite
systems may reach its limits. A case where we treated correlation
effects in TF approximation was, as already mentioned, the level
density parameter \cite{R02}. Pairing correlations in finite nuclei
have also already successfully been treated in the past \cite{R05}.
This is one of the aspects which we shall consider again in this work
in more detail. 

In this work we want to dwell on an aspect of Thomas-Fermi theory
which in the past has been exploited only very little. This concerns
the evaluation of matrix elements averaged over a certain energy
interval which may be typically of the order of $\hbar \omega$, i.e.,
the separation of major shells. 
 It must be pointed out that this is, to our knowledge, the first
attempt to evaluate not only one-body but also two-body matrix
elements in the
TF approximation. This semiclassical calculation provides the smoothly
varying part of the matrix elements dropping the shell effects
according to the idea of the Strutinsky averaging method \cite{R08}.
We will describe several tests of the accuracy of the TF method for
on-shell densities. In a first part we will develop our approach for
the matrix elements of single-particle operators, for given parity and
angular momentum. This goes along similar lines already developed in
the domain of systems with chaotic behaviour \cite{R21,R22,R23,R24}.
In a second part we will address our main objective, which is to show
that the method also works for two-body matrix elements. Some
preliminary results have been published previously in Ref. \cite{R06}.
As a specific example we will treat the pairing matrix elements.

Let us give a short summary of the approach we are going to develop.
Consider for example the expectation value of a single particle
operator $\hat{\it{O}}$ in some shell model state $| \nu \rangle$:
\begin{equation}
O_\nu = \langle \nu | \hat{O} | \nu \rangle =
Tr [ \hat{O}  | \nu \rangle \langle \nu | ].
\label{eq1}\end{equation}
Instead of knowing $O_\nu$ quantum state by quantum state it may some
times be advantageous and instructive to only know how the matrix
element (\ref{eq1}) changes as a function of energy. We therefore introduce a
single-particle matrix element averaged over the energy shell:
\begin{equation}
O(E) = Tr [ \hat{O} \hat{\rho}_E ] ,
\label{eq2} \end{equation}
where we call $\hat{\rho}_E$ the density matrix on the energy shell.
It is related with the so-called spectral density matrix and will be
defined immediately below.

The spectral density matrix $\delta (E -\hat{H})$ has the
characteristic discontinuous behaviour due to the quantization of the
eigenvalues of the single-particle Hamiltonian $\hat{H}$. It can be
written, however, as a sum of a smooth part $\tilde \delta (E
-\hat{H})$ and of a strongly oscillating part, i.e., $\delta (E
-\hat{H})= \tilde \delta (E-\hat{H}) + \delta_{\rm osc} (E-\hat{H})$
\cite{R1,R44,R6}. Analogously, the single-particle level density
$g(E)= Tr [\delta (E - \hat{H})]= \sum_\nu \delta (E -
\varepsilon_\nu)$
is obtained as a sum of two terms: $g(E)= \tilde g(E)+ g_{\rm osc}
(E)$, where $\tilde g$ and $g_{\rm osc}$ stand for the smooth and the
rapidly fluctuating contributions, respectively.
Using the smooth $\tilde \delta (E -\hat{H})$ and $\tilde g(E)$, we
define the density matrix averaged on the energy shell as
\begin{equation}
\hat{\rho}_E = \frac{1}{\tilde g(E)} \tilde \delta (E -\hat{H})
= \frac{1}{\tilde g(E)} \sum_\nu \tilde \delta (E - \varepsilon_\nu)
| \nu \rangle \langle \nu |.
\label{eq3} \end{equation}
It is therefore a smooth function of $E$. The smeared level density
$\tilde g(E)$ (per spin and isospin in this paper) in the denominator
of expression (\ref{eq3}) ensures the right normalization of
$\hat{\rho}_E$, since $\tilde g(E)= Tr [\tilde \delta (E - \hat{H})]$.

The smooth quantities entering eq.(\ref{eq3}) are to be evaluated in
some continuum limit \cite{R44,R6}: this is the case for example when
one introduces the Strutinsky averaging procedure \cite{R04,R08} or,
alternatively, and this is the approach we adopt in this work, it can
be done by replacing $\hat{H}$, the independent-particle (mean field)
Hamiltonian, by its classical counterpart $H_{cl}$ which corresponds
to the TF approximation \cite{R1}. Such an approximation has been used
very early by Migdal \cite{R2} and later, as already mentioned, in the
context of chaotic motion dynamics \cite{R21,R22,R23,R24}. Recently,
we have employed it to describe Bose condensates in traps \cite{R3},
but we are not aware of any systematic use in the context of nuclear
physics. The approach is not limited to the evaluation of expectation
values of single particle operators. Also the average behavior of
two-body matrix elements can be calculated. For instance the
semiclassical evaluation of the average pairing matrix element
\begin{equation}
v(E,E') = \frac{1}{\tilde g(E) \tilde g(E')} \sum_{\nu,\nu'} \tilde
\delta (E - \varepsilon_\nu) \, \tilde \delta(E' - \varepsilon_{\nu'}) \,
{\langle \Phi(\nu, \bar{\nu}) | v | \Phi(\nu', \bar{\nu}') \rangle} ,
\label{eq4} \end{equation}
where $|\Phi(\nu, \bar{\nu})\rangle$ is an {\em antisymmetric}
normalized two-body state constructed out of a state $| \nu \rangle$
and its time-reversed state $| \bar{\nu} \rangle$, can be of great
practical interest and shall be considered in this work. As it is
known \cite{R1,R01}, the Strutinsky method averages the density matrix
over an energy interval corresponding roughly to the distance between
two major shells. Implicitly the same holds if the equivalent
Wigner-Kirkwood expansion (TF approximation at lowest order) is used
for $\hat{\rho}_E$.

In detail our paper is organized as follows. In the next section the
Wigner function on the energy shell is introduced and applied to the
evaluation of some semiclassical one-body and two-body matrix elements
of physical interest. Our conclusions are given in the last section.

\section{Wigner function on the Energy Shell}

As stated in eq. (\ref{eq3}) of the introduction, we are interested
in  the density matrix $\hat{\rho}_E$ on the energy shell.
Consider the Wigner transform \cite{R1} of the density matrix
(\ref{eq3}),  namely
\begin{equation}
f_E(\mbox{\boldmath$R$},\mbox{\boldmath$p$}) = 
\int d \mbox{\boldmath$s$} e^{- i \mbox{\boldmath$p$} \mbox{\boldmath$s$}
/\hbar} \langle \mbox{\boldmath$R$} + \mbox{\boldmath$s$}/2
| \hat\rho_E | \mbox{\boldmath$R$} - \mbox{\boldmath$s$}/2 \rangle,
\label{eq4a} \end{equation}
where 
$$\mbox{\boldmath$R$}=(\mbox{\boldmath$r$} + \mbox{\boldmath$r$}')/2 ,
\qquad  \mbox{\boldmath$s$}=\mbox{\boldmath$r$}-\mbox{\boldmath$r$}'$$
are the centroid and relative coordinates respectively. In order to 
obtain the TF approximation plus $\hbar$ corrections to the Wigner
function on the energy shell (\ref{eq4a}), it is convenient to
differentiate with respect to $E$  the Wigner-Kirkwood expansion
of the full single-particle
one-body density matrix $\hat{\rho}= \Theta (E - \hat{H})$, which is
amply given in the literature \cite{R1}. Up to order $\hbar^2$ the
result is
\begin{eqnarray}
f_E(\mbox{\boldmath$R$},\mbox{\boldmath$p$}) &=& \frac{1}{\tilde g(E)}
\bigg[ \delta(E -H_{cl}) -
\frac{\hbar^2}{8M} \mbox{\boldmath$\nabla$}^2 V \delta''(E -H_{cl})
\nonumber \\
&&+
\frac{\hbar^2}{24M}\big[
(\mbox{\boldmath$\nabla$} V)^2 + \frac{1}{M} (\mbox{\boldmath$p$}
\cdot \mbox{\boldmath$\nabla$})^2 V \big]
\delta'''(E - H_{cl}) + {\mathcal O} (\hbar^4) \bigg].
\label {eq5} \end{eqnarray}
One should realize that $\tilde g(E)$ also contains $\hbar$ corrections and
that, strictly speaking, in order to get a consistent expansion of
$f_E$ one should also take into account the $\hbar$ expansion of
$\tilde g(E)$ and then correctly sort out relation (\ref{eq5}) to order
$\hbar^2$ (see also comments at the end of Section II.A).

The first term in eq. (\ref{eq5}) represents evidently the pure TF
approximation which is of lowest order in $\hbar$. In a first attempt and
to assess the accuracy of our approximation we will content ourselves
with  the TF approximation. Integration over the momenta yields the
local  density on the energy shell:
\begin{equation}
\rho^{TF}_E (\mbox{\boldmath$R$}) = 
\frac{1}{(2 \pi \hbar)^3} \int
d \mbox{\boldmath$p$} f^{TF}_E
(\mbox{\boldmath$R$},\mbox{\boldmath$p$}) =
\frac{M k_E(\mbox{\boldmath$R$})}{2 \pi^2 \hbar^2 \tilde g(E)},
\label{eq6} \end{equation}
where
\begin{equation}
k_E(\mbox{\boldmath$R$}) = \sqrt{\frac{2M}{\hbar^2}
(E - V(\mbox{\boldmath$R$}))}
\label{eq7} \end{equation}
is the local momentum at the energy $E$ in the potential $V(
\mbox{\boldmath$R$})$ and the level density $\tilde g(E)$ is given by
\begin{equation}
\tilde g(E) = \int d \mbox{\boldmath$R$}
\frac{M k_E(\mbox{\boldmath$R$})}{2 \pi^2 \hbar^2 }.
\label{eq7a} \end{equation}
 For the
following it is important to first
elaborate on the meaning and accuracy of this density on the energy
shell. For demonstration purposes we will take as an example the
spherical HO potential but later we will see
that our method works equally well for a Woods-Saxon (WS) potential.

In fig. 1 we display the quantal (solid line) and TF (dash-dotted
line) densities of the $N=4$
and $N=5$ HO shells with $\hbar \omega = 41 A^{-1/3}$
and $A=224$. For the TF densities we have taken the quantal energies.
We see that in both cases the TF result passes accurately through the
average, terminating at the classical turning point defined by
\begin{equation}
E = V(\mbox{\boldmath$R$}_{cl}).
\label{eq23b} \end{equation}    
The features of the TF densities on the energy shell are quite analogous 
to the ones already known for the full TF density \cite{R1}. However, on 
the quantal side one remarks that there is a
strong difference between odd parity ($N=5$) and even parity
($N=4$) shells. The former shows a pronounced hole at the origin
whereas the second ones shows, on the contrary, an enhancement. Both
features can obviously be related to the absence or presence of
the $s$-wave contributions in the corresponding HO
shell, respectively. One may try to recover
this even-odd parity effect in projecting the TF density matrix
on good parity. This is easily done as follows. We calculate the
inverse Wigner transform  of
$f^{TF}_E(\mbox{\boldmath$R$},\mbox{\boldmath$p$})$. This yields
\begin{equation}
\rho^{TF}_E(\mbox{\boldmath$r$},\mbox{\boldmath$r$}')=
\rho^{TF}_E(\mbox{\boldmath$R$})
j_0[s k_E(\mbox{\boldmath$R$})] ,
\label{eq8} \end{equation}
where $j_0$ is the zeroth-order spherical Bessel function. Now the
even/odd parity density on the energy shell is obtained as
\begin{eqnarray}
\rho^{e/o}_E(\mbox{\boldmath$r$}) = \frac{1}{2} \bigg[
\rho^{TF}_E(\mbox{\boldmath$r$},\mbox{\boldmath$r$}')
\pm  \rho^{TF}_E(\mbox{\boldmath$r$},-
\mbox{\boldmath$r$}') \bigg]_{\mbox{\boldmath$r$}=
\mbox{\boldmath$r$}'} =
\frac{1}{2} \bigg[ \rho^{TF}_E(\mbox{\boldmath$r$}) \pm \rho^{TF}_E (0)
j_0[2 r k_E(0)] \bigg].
\label{eq9} \end{eqnarray}
We have drawn this expression in fig. 1 (dashed lines) as well. The
bump (hole) structure exhibited by the quantal density is now well
reproduced in the interior. The agreement only deteriorates near the
classical turning point. One should mention, however, that in spite of
the seemingly rather spectacular improvement of formula (\ref{eq9})
over (\ref{eq6}), the former presents some small problems. This
concerns the behavior of (\ref{eq9}) around the turning point. The
presence of the second term in (\ref{eq9}) can induce a slightly
negative value of the density around the turning point. Also the
second term is not naturally limited to $r$ values inside the turning
point and thus is oscillating around zero due to the Bessel function.
This leads to ambiguities in evaluating matrix elements such as
(\ref{eq2}) which, however, numerically are rather unimportant. Thus
we advocate to use (\ref{eq6}) instead of (\ref{eq9}), excepting for
some problems where the even/odd bump structure may be particularly
important. The latter may for example be the case for the evaluation
of matrix elements of operators more concentrated at the nuclear
interior like e.g. $1/r^2$, etc.

\subsection{One-body matrix elements}
We now proceed to calculate in TF approximation as a function of the
energy the rms radius of a nucleon confined in a WS potential
with $V_0=- 44$ MeV, $a$= 0.67 fm and $R=1.27 A^{1/3}$ fm with $A$=224
nucleons. We choose the rms radius for demonstration purposes but we
could have taken as well any other smoothly varying single-particle
operator. We use the TF approximation (\ref{eq8}) and show the results
(dashed line) together with their quantum mechanical counterparts,
represented by dots, in fig. 2. We see that the TF calculation very
nicely passes through the average of the scattered quantal values, with 
the exception of the lowest $s$-state. This is a first confirmation of the
accuracy of our approach.

In a next step we want to project the TF density matrix on different
partial waves and calculate matrix elements as a function of the
energy for different $l$-values. One way to project on partial waves
has been elaborated by Hasse \cite{R4}. There one pre-multiplies the
Wigner function with the semiclassical projectors on the orbital
angular momentum and its $z$-component, i.e.,
\begin{equation}
f_{E,l,m}(\mbox{\boldmath$R$},\mbox{\boldmath$p$}) =
\frac{1}{\tilde g_{E,l,m}}
\delta(l - |\mbox{\boldmath$R$} \times \mbox{\boldmath$p$}|)
\delta(m - l_z(R,p))
\delta(E - H_{cl}) .
\label{eq13} \end{equation}
Then one can calculate single particle matrix elements as
\begin{equation}
{\cal{O}}(E,l,m) = \int \frac{d \mbox{\boldmath$R$}
d \mbox{\boldmath$p$}}{(2 \pi \hbar)^3}
{\cal{O}}(\mbox{\boldmath$R$},\mbox{\boldmath$p$})
f_{E,l,m} (\mbox{\boldmath$R$},\mbox{\boldmath$p$})
\label{eq14} \end{equation}
For local operators it is sufficient to know the density, which can be
obtained in integrating (\ref{eq13}) over momenta. Assuming spherical
symmetry we can also sum over the $m$ quantum numbers and after some
algebra one finally finds \cite{R4}:
\begin{equation} \rho_{E,l}(\mbox{\boldmath$R$}) =
\frac{2 l + 1}{8 \pi^2 R^2 \tilde g_{E,l}} \sqrt{\frac{2M}{\hbar^2}}
\left[ E -V(R) - \frac{(l + \frac{1}{2})^2 \hbar^2}{2 M
R^2} \right]^{-1/2} \Theta \left( E -V(R) -
\frac{(l + \frac{1}{2})^2 \hbar^2}{2 M R^2} \right) . 
\label{eq15} \end{equation}
where the level density $\tilde g_{E,l}$ is chosen in such a way that the 
integral of (\ref{eq15}) over the available $R$ space is normalized
to unity. Let us mention that we have
replaced $l(l+1)$ by $(l + \frac{1}{2})^2$ as it is done in the WKB
method to recover the right asymptotic behaviour of the wave function
in the free ($V(R)=0$) case \cite{R43}. 

We recently employed, however, a different way to do the
$l$-projection which for some purposes may be more convenient \cite{R41}.
For this we first perform the inverse Wigner-transform of the TF part of
eq. (\ref{eq5}):
\begin{equation}
\rho_E(\mbox{\boldmath$r$},\mbox{\boldmath$r$}') =
\frac{1}{\tilde g(E)} \int \frac{d \mbox{\boldmath$p$}}
{(2 \pi \hbar)^3}
e^{i \mbox{\boldmath$\scriptstyle p$}
\cdot\mbox{\boldmath$\scriptstyle r$}/\hbar}
e^{-i\mbox{\boldmath$\scriptstyle p$}
\cdot\mbox{\boldmath$\scriptstyle r$}'/\hbar}
\delta(E - \frac{p^2}{2 M} - V(\mbox{\boldmath$R$})).
\label{eq16} \end{equation}
Expanding the plane waves in spherical harmonics,
\begin{equation}
e^{i \mbox{\boldmath$\scriptstyle k$}
\cdot\mbox{\boldmath$\scriptstyle r$}}
= 4 \pi
\sum_{l,m} (-i)^l j_l(k r) Y_{l m}(\Omega_k) Y_{l m}^{*}(\Omega_r),
\label{eq17} \end{equation}
we can read off the $l$-projected density matrix. For the local density we
then obtain
\begin{equation}
\rho_{E,l,m}(\mbox{\boldmath$R$}) =
\frac{1}{\tilde g_{E,l,m}} \frac{2M}{\pi \hbar^2}
k_E(R) \bigg[j_l(R k_E(R)) \bigg]^2 |Y_{l m}(\Omega)|^2
\, \Theta ( E - V(\mbox{\boldmath$R$}) ) ,
\label{eq18} \end{equation}
where $\tilde g_{E,l,m}$ is again chosen to ensure the right normalization 
of (\ref{eq18}), i.e., $\int d \mbox{\boldmath$R$} 
\rho_{E,l,m}(\mbox{\boldmath$R$})= 1$.
We can calculate for example the mean square radius
as a function of $E$ for different $l$ values:
\begin{equation}
\langle R^2 \rangle_{E,l} =
\int d \mbox{\boldmath$R$} R^2 \rho_{E,l,m}(\mbox{\boldmath$R$}) .
\label{eq19} \end{equation}
Notice that eq.(\ref{eq19}) becomes independent of $m$ after the
angular integration. 

In table I we show various moments ${\langle R^n
\rangle}^{1/n}$ (in fm) obtained using the TF local densities on the
energy shell provided by eqs.(\ref{eq15}) and (\ref{eq18}), as
compared with the corresponding quantal values for the aforementioned
WS potential with $A=224$ nucleons. From these tables one can see that
the quantal rms radius ($n=2$) of each $nl$ state is, on average, well
reproduced using the semiclassical $l$-projected on-shell densities
given by the TF approaches eqs.(\ref{eq15}) and (\ref{eq18}), where
$E$ has been replaced by the quantal eigenvalue corresponding to the
$nl$ state (for consistency we should perform a WKB quantization from
which we refrain for simplicity as it should not affect the result by
much). A more quantitative analysis shows that the quantal rms radii
are reproduced by eq.(\ref{eq15}) within 2\% in average over the range
of energies considered, while using eq.(\ref{eq18}) the average
relative error is around 4\%. Higher moments obtained with the TF
densities on the energy shell also reproduce reasonably well the
results of the full quantal calculation. For the highest moment
considered here ($n=10$), the average relative error with respect to
the quantal calculation is around 4.3\% for both semiclassical
approximations. One should point out that the TF local densities on
the energy shell are free of the shell effects that are present in the
quantal calculation. Actually, the TF results represent the shell
averaged values of the moments and their difference with the quantal
calculations provides an estimate of the shell correction for the
considered state. Of course, a precise calculation of the shell
correction in the considered moments would require a Strutinsky
calculation which is not an easy task for a WS potential. For a
related discussion we refer the reader to \cite{R42}, where the
moments ranging from $n=1$ to $n=10$ of the total density of $A=224$
particles in a HO potential were evaluated using several semiclassical
approaches and the Strutinsky averaging method. The fact that TF works
for moments as high as $n=10$ makes us believe that one can certainly
consider in our approach all single-particle operators which are
low-order polynomials in the phase-space variables.

 It is also instructive to directly compare the densities. For this 
we again take $E$ equal to its quantal value in eqs.(\ref{eq15}) and
(\ref{eq18}). The comparison between the quantal and TF on-shell
densities is shown in figs. 3-5. In each one of these figures the
quantal on-shell density (solid line) and the semiclassical densities
provided by eq.(\ref{eq15}) (dash-dotted line) and eq.(\ref{eq18})
(dashed line) are displayed for all the bound $s$ and $p$ states of a
WS potential with $A=224$. From these figures one can see
that the quantal on-shell densities are rather well reproduced by the
TF approach eq.(\ref{eq18}). In particular the quantal $1l$ on-shell
densities are well reproduced by eq.(\ref{eq18}) in the interior.
However, as it happens in fig. 1, there are discrepancies between the
quantal and TF densities in the outer part around the classical
turning point. For $nl$ states ($n > 1$), the inner $n-1$ bumps are
well reproduced by eq.(\ref{eq18}) and the agreement deteriorates in
the outer bumps due to the classical turning point where the TF
densities vanish. On the other hand the TF on-shell densities obtained
with eq.(\ref{eq15}) do not reproduce the quantal density profiles at
all. As it can be seen from eq.(\ref{eq15}), this density is defined
in  the region in between the two roots (turning points) of
\begin{equation}
E -V(R) - \frac{(l + \frac{1}{2})^2 \hbar^2}{2 M R^2} = 0 .
\label{eq20a} \end{equation}
In this approach the three-dimensional problem has been reduced to an
equivalent one-dimensional problem for $R$ with an effective potential
that in addition to $V(R)$ contains the centrifugal barrier, as it
happens in the WKB method. Thus, in this case we find two different
turning points. The largest root of (\ref{eq20a}) is very close 
to the classical turning point of eq.(\ref{eq18})
 given by $k_E=0$ [see eq.(\ref{eq7})], while the
smallest root gives the inner turning point due to the centrifugal
barrier. Since the TF on-shell density (\ref{eq15}) has square root
singularities at the two turning points, its integral as well as the
corresponding expectation values converge.

We arrive at the, at first sight, paradoxical result that the
densities (\ref{eq15}) which have no detailed resemblance with the
quantal ones reproduce the rms values (and very likely most of other
expectation values of smoothly varying operators) better than the
densities given in (\ref{eq18}), which show quite reasonable overall
behaviour in comparison with the quantal results. We here find an
illustrating example that the Thomas-Fermi and Wigner-Kirkwood local
densities are to be regarded as mathematical distributions, in the
sense that in spite of their possible divergences they yield finite,
accurate results when used to  compute a restricted class of expectation 
values by
integration \cite{R44,R42}. In this respect we again refer the reader
to tables I and II where he finds confirmation of our statement. A
similar situation is found in computing the kinetic energy for a
bosonic system in Ref. \cite{R3}: the semiclassical and quantal
kinetic energy densities are clearly different but give similar
values of the integrated kinetic energy. On the other hand, one should
note that sorting out correctly the various orders in $\hbar$ is very
important to achieve optimal results as shown on other occasions
\cite{R42,R46}. In expression (\ref{eq18}) there remain some $\hbar$
corrections to all orders in the form of the spherical harmonics which
are quantal wave functions, i.e., solutions of a Schr\"odinger
equation. This mixing of resummation in $\hbar$ on the one hand and of
lowest order in $\hbar$ on the other hand [in the form of the
$\delta$-function in (\ref{eq16})] finally makes the on-shell
density (\ref{eq18}) slightly less accurate than (\ref{eq15}),
which represents the correct $\hbar \to 0$ limit as shown in Ref.
\cite{R4}. 

\subsection{Two-body matrix elements}

As a further interesting application we want to consider the
semiclassical evaluation of average two-body matrix elements. An
example of particular interest is e.g. the case of matrix elements of
the pairing type $\langle \Phi(\nu, \bar{\nu}) | v | \Phi(\nu',
\bar{\nu}') \rangle$ [see eq.\ (\ref{eq4})], which we shall address in
this section for identical nucleons. It is straightforward to recast
the state-dependent pairing matrix element as
\begin{equation}
\langle \Phi(\nu, \bar{\nu}) | v | \Phi(\nu', \bar{\nu}') \rangle =
\langle \nu \bar{\nu} |v | \nu' \bar{\nu}' \rangle -
\langle \nu \bar{\nu} |v | \bar{\nu}' \nu' \rangle .
\label{eq_pair} \end{equation}
The two-particle states $|\nu \bar{\nu} \rangle$ on the r.h.s.\ are
product states of $| \nu \rangle$ and $|\bar{\nu} \rangle$. The
states $| \nu \rangle$ are represented by single-particle wave
functions $\phi_\nu(\mbox{\boldmath$r$},\sigma)=
\phi_{nlm}(\mbox{\boldmath$r$}) \, \chi_\sigma$ (with $\sigma= \pm
\frac{1}{2}$). Assuming spherical symmetry, and considering that the
time reversal
of $| \nu \rangle$ involves $\hat{T} (Y_{lm} \, \chi_\sigma) =
(-1)^m Y_{l,-m} \, (-1)^{1/2-\sigma} \chi_{-\sigma} =
(-1)^{1/2-\sigma} Y_{lm}^* \, \chi_{-\sigma}$, one finds
\begin{equation}
\frac{1}{4}
\sum_{m,m'} \sum_{\sigma,\sigma'} \,
\langle \Phi(\nu, \bar{\nu}) | v | \Phi(\nu', \bar{\nu}') \rangle =
\sum_{m,m'} 
\int d \mbox{\boldmath$r$} d \mbox{\boldmath$r$}' \,
\phi_{nlm} (\mbox{\boldmath$r$}') 
\phi_{nlm}^* (\mbox{\boldmath$r$}) 
\, v(\mbox{\boldmath$r$} - \mbox{\boldmath$r$}') \,
\phi_{n'l'm'} (\mbox{\boldmath$r$}) 
\phi_{n'l'm'}^* (\mbox{\boldmath$r$}') .
\label{eqxx0} \end{equation}
According to this result, we obtain the following expression for the
average pairing matrix elements of eq.(\ref{eq4}):
\begin{equation}
v(E,E') = 
\int d \mbox{\boldmath$r$} d \mbox{\boldmath$r$}'
\rho_E (\mbox{\boldmath$r$},\mbox{\boldmath$r$}')
v(\mbox{\boldmath$r$} - \mbox{\boldmath$r$}')
\rho_{E'} (\mbox{\boldmath$r$}',\mbox{\boldmath$r$}) ,
\label{eqxx1} \end{equation}
where $\rho_E (\mbox{\boldmath$r$},\mbox{\boldmath$r$}')
= \langle  \mbox{\boldmath$r$} | \hat\rho_E | \mbox{\boldmath$r$}'
\rangle$. 
In TF approximation the non-local on-shell density matrix
$\rho_E (\mbox{\boldmath$r$},\mbox{\boldmath$r$}')$
is given by eq.(\ref{eq8}). We see that it is a symmetric function in
$\mbox{\boldmath$r$}$ and $\mbox{\boldmath$r$}'$.

For the case of a force
$v_0 \delta (\mbox{\boldmath$r$} - \mbox{\boldmath$r'$})$
eq.(\ref{eqxx1}) reduces to (a practically identical expression can be
found in Ref. \cite{Tondeur})
\begin{equation}
v(E,E') = v_0 \int d \mbox{\boldmath$r$}
\rho_E(\mbox{\boldmath$r$}) \rho_{E'}(\mbox{\boldmath$r$}) .
\label{eq11} \end{equation}
Using the TF expression (\ref{eq6}) for $\rho_E(\mbox{\boldmath$r$})$
we can evaluate (\ref{eq11}) with a HO potential
$V(\mbox{\boldmath$r$})= m \omega_0^2 r^2/2$ and compare with the
quantum mechanical matrix elements averaged on each major shell $N$ of
energy $E= E_N= (N+\frac{3}{2}) \hbar \omega$. This is done in Table
II with $v_0= -345.723$ MeV fm$^3$ and $\hbar \omega_0= 41A^{-1/3}$
MeV. We again see that the semiclassical results agree very well with
the averaged quantal values, even for the non-diagonal elements.

With the above positive experience at hand, next we proceed to
calculate the average pairing matrix elements
$v(\varepsilon_F,\varepsilon_F)$ of the Gogny D1S force \cite{R5}
which is known to reproduce experimental gap values when used in
microscopic Hartree-Fock-Bogolyubov calculations \cite{R51}. Writing
the diagonal matrix element (\ref{eq4}) at the Fermi energy 
$\varepsilon_F$, by using (\ref{eqxx1}),
and expressing it through the lowest-order Wigner function in
inverting (\ref{eq4a}), one arrives in TF approximation at
\begin{equation}
v(\varepsilon_F,\varepsilon_F) = \frac{1}{{\tilde g(\varepsilon_F)}^2} \int d
\mbox{\boldmath$R$}
\int \frac{d \mbox{\boldmath$p$} d \mbox{\boldmath$p$}'}{(2 \pi \hbar)^6} \delta(\varepsilon_F
- H_{cl}(\mbox{\boldmath$R$},\mbox{\boldmath$p$}))
v(\mbox{\boldmath$p$}-\mbox{\boldmath$p$}')
\delta(\varepsilon_F-H_{cl}(\mbox{\boldmath$R$},\mbox{\boldmath$p$}'))
. \label{eq12} \end{equation}
Here, $H_{cl}= p^2/2{M^*}+V(\mbox{\boldmath$R$})$ is the classical
 Hamiltonian of independent particles with effective mass $M^*$ (see
below) moving in an external potential well
$V(\mbox{\boldmath$R$})$, and
$v(\mbox{\boldmath$p$}-\mbox{\boldmath$p$}')$
is the Fourier transform of the particle-particle part of the Gogny
force which describes the pairing. For the numerical application we
use for $V(\mbox{\boldmath$R$})$ a slight variant of the potential
given by Shlomo \cite{R6}: 
\begin{equation}
V(R) = - \frac{V_0}{1 + \exp{(-R_0/d)}}
+ \frac{V_0}{1 + \exp{((R-R_0)/d)}}
\label{eq21} \end{equation}
with
\begin{equation}
R_0 = \frac{1.12 A^{1/3}+1}{[1+(\pi d/R_0)^2]^{1/3}} \mbox{ fm},
\qquad \qquad d=0.70 \mbox{ fm}, \qquad {\rm and} \qquad V_0=-54
\mbox{ MeV}.
\label{eq21a} \end{equation}
In this equation $R_0$ has to be determined iteratively.

 Equation (\ref{eq12}) can be reduced to a one dimensional integral over
$R$ which can be performed numerically:
\begin{equation}
v(\varepsilon_F,\varepsilon_F) =  \sum_{c=1}^2 z_c \frac{1}{\mu_c^2}
\int_0^{R_c} dR R^2 B
\exp{[-a(\varepsilon_F - V(R))]}
\sinh{[a(\varepsilon_F - V(R))]} ,
\label{eq23} \end{equation}
where $R_c$ is the classical turning point defined in (\ref{eq23b}) and
\begin{equation}
B = \frac{1}{4 \pi^3 \tilde g(\varepsilon_F)^2}
\left(\frac{2M^*}{\hbar^2}\right)^2
, \qquad a= \frac{M^* \mu_c^2}{\hbar^2}, \qquad
{\rm and} \qquad  z_c = \pi^{3/2} \mu_c^3 (W_c - B_c - H_c + M_c).
\label{eq23a} \end{equation}
The factors $z_c$ correspond to pairing in the $S=0$ and $T=1$
channel and are written in terms of the parameters of the Gogny
force $W_c$, $B_c$, $H_c$, $M_c$ and $\mu_c$ \cite{R5}. 
We have introduced the position-dependent effective mass
$M^*(\mbox{\boldmath$R$})$ from the Gogny force in order to make the
calculation of the pairing matrix element more realistic. It is
obtained by evaluating at 
$k= k_{\varepsilon_F}(\mbox{\boldmath$R$})$ the position- and
momentum-dependent effective mass \cite{R61}
\begin{equation}
\frac{M}{M^*(\mbox{\boldmath$R$},k)} =
1+ \frac{M}{\hbar^2 k}
\frac{\partial}{\partial k} U(\mbox{\boldmath$R$},k) ,
\label{eq23c} \end{equation}
where $U(\mbox{\boldmath$R$},k)$ is the Wigner transform of the
single-particle potential obtained from the Gogny interaction assumed
spherically symmetric in $\mbox{\boldmath$k$}$.

 Also the level density $\tilde g(\varepsilon_F)$
is calculated in the TF approach using the same potential and effective mass:
\begin{equation}
\tilde g(\varepsilon_F) = \frac{1}{\pi} \int_0^{R_c} dR R^2
\left(\frac{2M^*(\mbox{\boldmath$R$})}{\hbar^2}\right)^{3/2} 
\sqrt{\varepsilon_F - V(\mbox{\boldmath$R$})}.
\label{eq24} \end{equation}

In fig. 6 we show $A v(\varepsilon_F,\varepsilon_F)$ as a function of
the mass number $A$. The Coulomb force has been switched off in the
present calculation. We see that there is a quite pronounced $A$
dependence which is somewhat in contradiction with the value $G \sim
28/A$ MeV for the constant pairing matrix element at the Fermi level
usually employed in more schematic treatments of the nuclear pairing.
On the other hand, if we calculate $A v(\varepsilon_F,\varepsilon_F)$ not
with a WS potential but with the HO potential, we obtain a practically
constant value. This may indicate that the $A^{-1}$ dependence of the
constant pairing matrix element is better fulfilled in conjunction
with a harmonic potential. The difference in the behaviour with
$A$ using the HO and WS potentials may come from the absence (HO) or
presence (WS) of a surface contribution to
$v(\varepsilon_F,\varepsilon_F)$, like it is the case for the level
density \cite{R1}. 
%

\section{Conclusions}
In this work we showed how average nuclear one- and two-body matrix
elements can very efficiently be evaluated using the Thomas-Fermi
approach. The main ingredient is to replace the density matrix for a
given quantum state $\hat{\rho}_\nu = |\nu \rangle \langle \nu |$ by
its counterpart averaged over the energy shell
$$
\hat{\rho}_E = \frac{1}{\tilde g(E)} \sum_\nu
\tilde \delta (E -\varepsilon_\nu) | \nu \rangle \langle \nu |
$$
with simultaneous application of the Wigner-Kirkwood $\hbar$ expansion
for the smoothly varying spectral density $\tilde \delta(E - \hat{H})$
and level density $\tilde g(E)= Tr [\tilde \delta (E - \hat{H})]$.

We calculated one- and two-body matrix elements restricting ourselves,
in this exploratory work, to the lowest order, i.e., the pure
Thomas-Fermi approximation. We compared quantal and semiclassical
values of the matrix elements using harmonic oscillator and
Woods-Saxon type of potentials. We did this also for parity projected
and for angular momentum projected Thomas-Fermi theory. In all the
cases close agreement with the average quantal behaviours was found,
showing the accuracy of the method. As in the case of the well
tested Wigner-Kirkwood expansion of the full density matrix \cite{R1},
one expects that some improvement also could be achieved for the
matrix elements by inclusion of $\hbar$ corrections.

With the positive result for the single-particle matrix elements at
hand, we also calculated the average pairing matrix elements of some
effective nuclear two-body forces. First, we used a delta interaction
and compared diagonal and off-diagonal semiclassical elements with the
corresponding quantal values. Again the TF values nicely reproduce the
quantum results on average. Next we estimated semiclassically the
diagonal pairing matrix elements of the Gogny D1S force at the Fermi
energy. Since the Gogny force is known to reproduce very well nuclear
pairing properties \cite{R51}, it is interesting to evaluate e.g. the
$A$ dependence of its pairing matrix elements around the Fermi energy
and to see to which extent the common assumption of a $A^{-1}$
dependence holds. Using for the mean field a potential of the
Woods-Saxon type, it turned out that the fall off of the pairing
matrix element is stronger than the $A^{-1}$ law. For this problem we
have no comparison with quantum values available, but the experience
with the one-body matrix elements and the pairing matrix elements for
the delta force, makes us believe that the values shown in fig. 6 are
also reasonably accurate. The stronger than $A^{-1}$ decrease observed
in fig. 6 has its origin very likely in the presence of a surface
contribution implicit from the use of a WS potential, whereas the use
of a HO potential with the absence of a surface shows agreement with
the $A^{-1}$ law.

It should be emphasized that to our knowledge the TF method to
calculate average matrix elements of a two-body force has never been
applied before.
We think that the conclusive study of this work will allow to use
average matrix elements for the calculation of many nuclear
quantities where fine shell effects are not needed such as optical
potentials, giant resonances and their widths, and many other
quantities where the average trend is of interest. In a future
publication we will show how the application of these techniques can
be used to study in a very transparent way the size dependence of the
average pairing gap in finite Fermi systems in an almost analytical
way.

\acknowledgments
P.S. wants to thank P. Leboeuf and N. Pavloff for useful discussions
and informations. Two of us (X.V. and M.C.) acknowledge financial
support from the DGI (Ministerio de Ciencia y Tecnolog{\'\i}a, Spain)
and FEDER under grant BFM2002-01868 and from DGR (Catalonia) under
grant 2001SGR-00064.

%


%
\begin{table}
\caption[]{Moments ${\langle R^n \rangle}^{1/n}$ (in fm) of several
$nl$ states for $A=224$ particles in the Woods-Saxon potential
described in the text. The full quantum mechanical values (QM) are
compared with those obtained with the semiclassical approaches
eqs.(\ref{eq15}) and (\ref{eq18}).}
\centering
\vspace{1.5cm}
\begin{tabular}{llll cc cc cc cc cc c}
 $nl$ && & & $E$ (MeV) & & $n=2$ & & $n=4$ & & $n=6$ & & $n=8$ & & $n=10$ 
\\
\hline
1s && QM & &  3.87 & & 4.021 & & 4.441 & & 4.780 & & 5.069 & & 5.323\\
&& eq.($\ref{eq15}$) & &       & & 4.040 & & 4.528 & & 4.824 & & 5.019 & & 5.160\\
&& eq.($\ref{eq18}$) & &       & & 4.299 & & 4.638 & & 4.863 & & 5.024 & & 5.143\\  
1p && QM & &  7.65 & & 4.732 & & 5.042 & & 5.307 & & 5.543 & & 5.758\\
&& eq.($\ref{eq15}$) & &       & & 4.653 & & 5.011 & & 5.267 & & 5.449 & & 5.583\\
&& eq.($\ref{eq18}$) & &       & & 4.900 & & 5.144 & & 5.321 & & 5.546 & & 5.561\\                                                                     
1d && QM & & 12.23 & & 5.226 & & 5.477 & & 5.701 & & 5.907 & & 6.099\\
&& eq.($\ref{eq15}$) & &       & & 5.123 & & 5.397 & & 5.612 & & 5.776 & & 5.900\\
&& eq.($\ref{eq18}$) & &       & & 5.323 & & 5.516 & & 5.664 & & 5.782 & & 5.876\\                                                                       
2s && QM & & 13.92 & & 4.746 & & 5.401 & & 5.780 & & 6.052 & & 6.275\\
&& eq.($\ref{eq15}$) & &       & & 4.668 & & 5.301 & & 5.662 & & 5.897 & & 6.064\\
&& eq.($\ref{eq18}$) & &       & & 4.683 & & 5.331 & & 5.679 & & 5.901 & & 6.060\\
1f && QM & & 17.49 & & 5.616 & & 5.831 & & 6.031 & & 6.218 & & 6.397\\
&& eq.($\ref{eq15}$) & &       & & 5.511 & & 5.731 & & 5.913 & & 6.057 & & 6.172\\
&& eq.($\ref{eq18}$) & &       & & 5.568 & & 5.819 & & 5.948 & & 6.054 & & 6.141\\                                                                    
2p && QM & & 20.08 & & 5.154 & & 5.771 & & 6.159 & & 6.439 & & 6.668\\
&& eq.($\ref{eq15}$) & &       & & 5.065 & & 5.653 & & 6.009 & & 6.244 & & 6.411\\
&& eq.($\ref{eq18}$) & &       & & 4.931 & & 5.601 & & 6.002 & & 6.256 & & 6.433\\
%
%
1g && QM & & 23.37 & & 5.948 & & 6.143 & & 6.327 & & 6.504 & & 6.679\\
&& eq.($\ref{eq15}$) & &       & & 5.836 & & 6.021 & & 6.179 & & 6.309 & & 6.414\\
&& eq.($\ref{eq18}$) & &       & & 5.942 & & 6.083 & & 6.200 & & 6.297 & & 6.379\\                                                                   
2d && QM & & 26.63 & & 5.587 & & 6.159 & & 6.547 & & 6.836 & & 7.078\\
&& eq.($\ref{eq15}$) & &       & & 5.480 & & 6.014 & & 6.354 & & 6.583 & & 6.747\\
&& eq.($\ref{eq18}$) & &       & & 5.293 & & 5.942 & & 6.356 & & 6.622 & & 6.802\\                                                                 
3s && QM & & 27.80 & & 5.485 & & 6.236 & & 6.679 & & 6.998 & & 7.263\\
&& eq.($\ref{eq15}$) & &       & & 5.374 & & 6.077 & & 6.467 & & 6.719 & & 6.895\\
&& eq.($\ref{eq18}$) & &       & & 5.156 & & 6.014 & & 6.483 & & 6.771 & & 6.963\\                                                                       
1h && QM & & 29.78 & & 6.252 & & 6.435 & & 6.612 & & 6.788 & & 6.967\\
&& eq.($\ref{eq15}$) & &       & & 6.128 & & 6.289 & & 6.429 & & 6.547 & & 6.644\\
&& eq.($\ref{eq18}$) & &       & & 6.177 & & 6.316 & & 6.427 & & 6.521 & & 6.601\\                                                                     
2f && QM & & 33.44 & & 6.063 & & 6.606 & & 6.999 & & 7.304 & & 7.584\\
&& eq.($\ref{eq15}$) & &       & & 5.923 & & 6.410 & & 6.731 & & 6.950 & & 7.108\\
&& eq.($\ref{eq18}$) & &       & & 5.775 & & 6.375 & & 6.766 & & 7.018 & & 7.188\\                                                                       
3p && QM & & 34.97 & & 6.097 & & 6.862 & & 7.341 & & 7.709 & & 8.037\\
&& eq.($\ref{eq15}$) & &       & & 5.939 & & 6.621 & & 7.004 & & 7.421 & & 7.421\\
&& eq.($\ref{eq18}$) & &       & & 5.866 & & 6.621 & & 7.036 & & 7.294 & & 7.469\\                                                                       
\end{tabular}
\label{tab1}\end{table}
\pagebreak
%
\begin{table}
\caption[]{Quantal (QM) and Thomas-Fermi (TF) averaged two-body matrix
elements (in MeV) of the $v(\mbox{\boldmath$r$},\mbox{\boldmath$r$}')
= -345.723 \, \delta(\mbox{\boldmath$r$} -
\mbox{\boldmath$r$}')$ force calculated with harmonic oscillator density
matrices on the energy shell for $A=224$ particles.}
\centering
\vspace{2cm}
\begin{tabular}{ll cc cc cc cc cc cc}
& & $N\backslash N'$ & 0 & & 1 & & 2 & & 3 & & 4 & & 5
\\
\hline
 QM & & 0 & $-$1.44 & & $-$0.72 & & $-$0.45 & & $-$0.32 & & $-$0.24 & & $-$0.19 \\
 TF & &   & $-$1.20 & & $-$0.68 & & $-$0.44 & & $-$0.31 & & $-$0.23 & & $-$0.18 \\
 QM & & 1 &         & & $-$0.60 & & $-$0.41 & & $-$0.29 & & $-$0.22 & & $-$0.18 \\
 TF & &   &         & & $-$0.56 & & $-$0.39 & & $-$0.29 & & $-$0.22 & & $-$0.18 \\
 QM & & 2 &         & &       & & $-$0.35 & & $-$0.27   & & $-$0.21 & & $-$0.17 \\
 TF & &   &         & &       & & $-$0.34 & & $-$0.26   & & $-$0.21 & & $-$0.17 \\
 QM & & 3 &         & &       & &         & & $-$0.24   & & $-$0.19 & & $-$0.16 \\
 TF & &   &         & &       & &         & & $-$0.23   & & $-$0.19 & & $-$0.16 \\
 QM & & 4 &         & &       & &         & &           & & $-$0.17 & & $-$0.15 \\
 TF & &   &         & &       & &         & &           & & $-$0.17 & & $-$0.15 \\
 QM & & 5 &         & &       & &         & &           & &         & & $-$0.14 \\
 TF & &   &         & &       & &         & &           & &         & & $-$0.13 \\
\end{tabular}
\label{tab2}\end{table}
\pagebreak

\section*{Figure captions}
\noindent
Figure 1.
 Quantal (solid line), pure TF (dashed-dotted line) and TF densities
projected on the good parity (dashed line) for the $N=4$ and
$N=5$ harmonic oscillator shells.\\
Figure 2.
Quantal and TF (dashed line) rms radii for a Woods-Saxon potential.\\
Figure 3.
1$s$ and 1$p$ on-shell densities in a Woods-Saxon potential calculated
quantally (solid line) and with the TF approximation using
eq. (\ref{eq15}) (dash-dotted line) and eq. (\ref{eq18}) (dashed line).
The dash-dotted vertical lines indicate the asymptotes of eq.
(\ref{eq15}).\\
Figure 4.
 Same as Figure 3 for the 2$s$ and 2$p$ on-shell densities.\\
Figure 5.
 Same as Figure 3 for the 3$s$ and 3$p$ on-shell densities.\\
Figure 6.
Two-body pairing matrix elements computed with the D1S Gogny force and
the Shlomo potential (\ref{eq21}) as a function of the mass number
$A$.


\begin{references}
%
\bibitem{R1}
 P. Ring and P. Schuck, {\it The Nuclear Many-Body Problem}
(Springer-Verlag, New York, 1980); M. Brack and R. K. Badhuri,
{\it Semiclassical Physics} (Addison-Wesley, Reading, MA, 1997).
%
\bibitem{R01}
M. Brack and H. C. Pauli, Nucl.\ Phys.\ {\bf A207}, 401 (1973);
B. K. Jennings, Nucl.\ Phys.\ {\bf A207}, 538 (1973).
%
%
\bibitem{R02}
R. W. Hasse and P. Schuck, Phys.\ Lett.\ B {\bf 179}, 313 (1986).
%
\bibitem{R03}
M. Durand, P. Schuck, and X. Vi\~nas, Z.\ Phys.\ {\bf A346}, 87
(1993); M. Centelles, X. Vi\~nas, and P. Schuck,
 Phys.\ Rev.\ C {\bf 53}, 1018 (1996).
%
\bibitem{R04}
R. K. Bhaduri and C. K. Ross,
Phys.\ Rev.\ Lett. {\bf 27}, 606 (1971);
B. K. Jennings and R. K. Bhaduri,
Nucl.\ Phys.\ A {\bf 237}, 149 (1975);
B. K. Jennings, R. K. Bhaduri, and M. Brack,
Phys.\ Rev.\ Lett. {\bf 34}, 228 (1975);
Nucl.\ Phys.\ A {\bf 253}, 29 (1975).
%
\bibitem{R05}
K. Taruishi and P. Schuck, Z. Phys.\ A {\bf 342}, 397 (1992);
P. Schuck and K. Taruishi, Phys.\ Lett.\ B {\bf 385}, 12 (1996).
%
%
\bibitem{R08}
M. Brack, J. Damg{\aa}rd, A. S. Jensen, H.C. Pauli, V. M. Strutinsky
and C. Y. Wong, Rev. \ Mod. \ Phys. \ {\bf 44} , 320 (1972).
%
\bibitem{R21} M. Wilkinson, J. Phys. A {\bf 20}, 2415
(1987).
%
\bibitem{R22}  B. Mehlig, Phys.\ Rev.\ E {\bf 59}, 390 (1999).
%
\bibitem{R23}  V. N. Kondratyev, Phys.\ Lett.\ A {\bf 179} 209,
(1993). 
%
\bibitem{R24} M. L. Du and J. B. Delos, Phys.\ Rev.\ A {\bf 38}, 1896
(1988); Phys.\ Rev.\ A {\bf 38}, 1913 (1988).
%
\bibitem{R06}
X. Vi\~nas, P. Schuck, M. Farine, M. Durand and M. Centelles,
Phys. of Atomic\ Nucl.\  {\bf 65}, 731 (2002); Yad.\ Fis. {\bf 65}, 764 
(2002).
%
\bibitem{R44} H. Krivine, M. Casas, and J. Martorell,
Ann. \ of \ Phys. \ (NY) {\bf 200}, 304 (1990).
%
\bibitem{R6}
 S. Shlomo, Nucl.\ Phys.\ {\bf A539}, 17 (1992).
%
%
\bibitem{R2}
 A. B. Migdal, Nucl.\ Phys.\ {\bf 13}, 655 (1959).
%
\bibitem{R3}
 P. Schuck and X. Vi\~nas, Phys.\ Rev.\ A {\bf 61}, 043603 (2000).
%
\bibitem{R4}
R. W. Hasse, Annales Universitatis Mariae Curie-Sklodowska Lublin
(Poland) {\bf XL/XLI} 191 (1965); Nucl.\ Phys.\ {\bf A467}, 407 (1987).
%
\bibitem{R43}
 L. D. Landau and E. M. Lifshitz, {\it Quantum Mechanics}
(Pergamon Press, New York, 1991).
%
\bibitem{R41} M. Urban, P. Schuck, and X. Vi\~nas, {\it cond-mat/0207261} 
(2002).
%
\bibitem{R42} M. Centelles, X. Vi\~nas, M. Durand, P. Schuck and D. 
Von-Eiff, Ann. \ of \ Phys. \ (NY) {\bf 266}, 207 (1998).
%
\bibitem{R46} P. Schuck and X. Vi\~nas,
 Phys.\ Lett.\ B {\bf 302}, 1 (1993).
%
\bibitem{Tondeur}
 F. Tondeur, Nucl.\ Phys.\ {\bf A315}, 353 (1979).
%
\bibitem{R5}
 J. F. Berger, M. Girod, and D. Gogny, Comput.\ Phys.\ Commun.\ {\bf
63}, 365 (1991); Nucl.\ Phys.\ {\bf A502}, 85c (1989).
%
\bibitem{R51}
M. Kleban, B. Nerlo-Pomorska, J. F. Berger, J. Decharg\'e, M. Girod,
and S. Hilaire, Phys.\ Rev.\ C {\bf 65}, 024309 (2002).
 %
\bibitem{R61}
 V. B. Soubbotin and X. Vi\~nas, Nucl.\ Phys.\ {\bf A665}, 291 (2000).
%
\end{references}
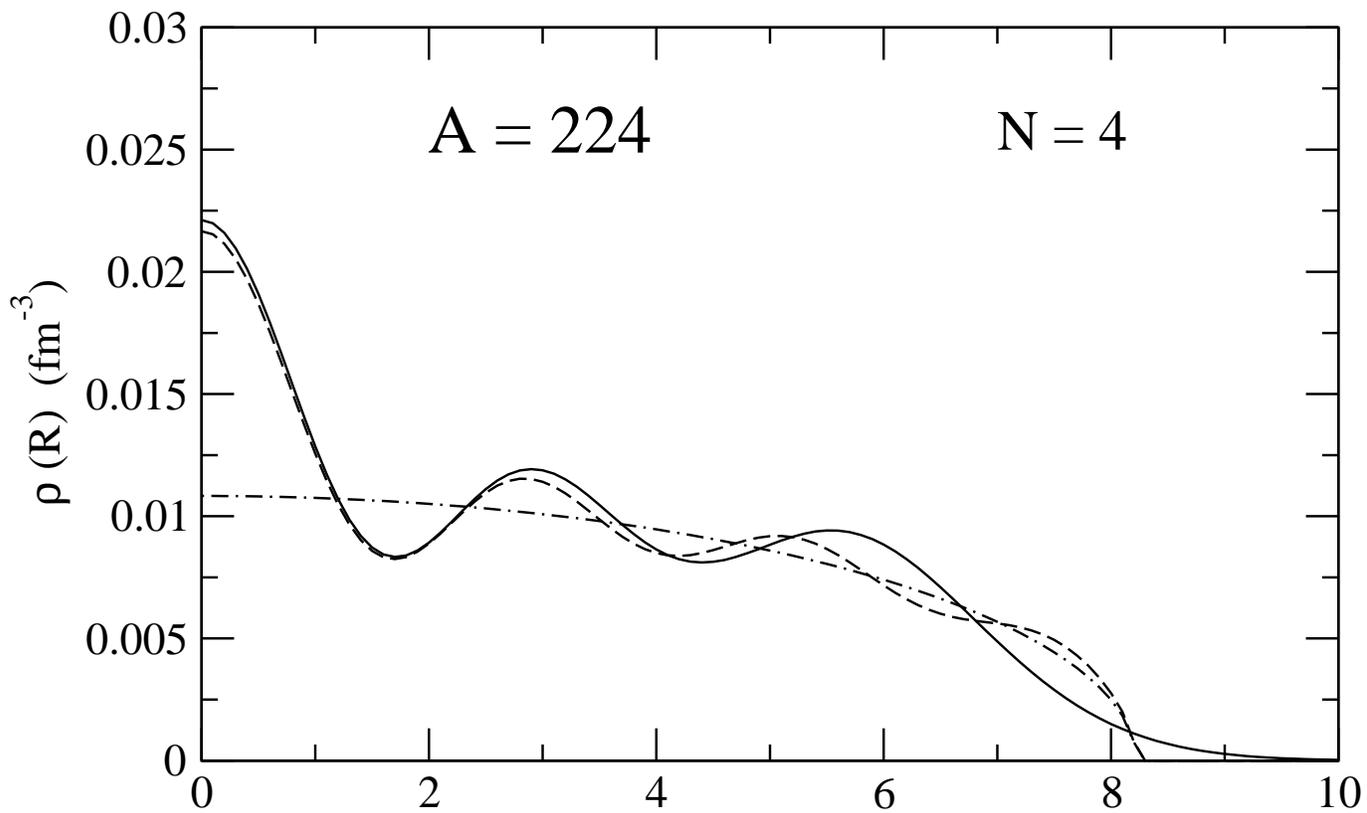
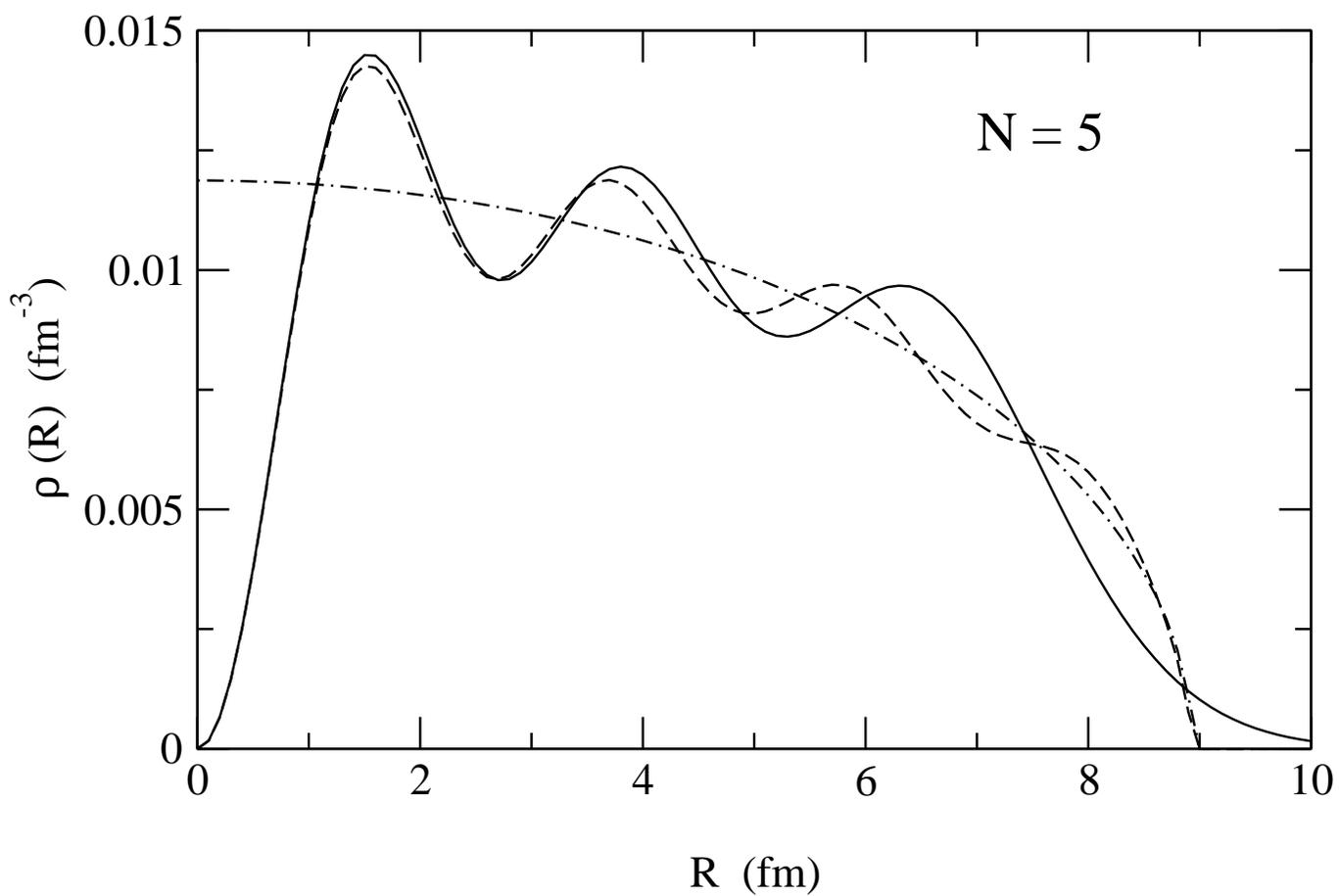
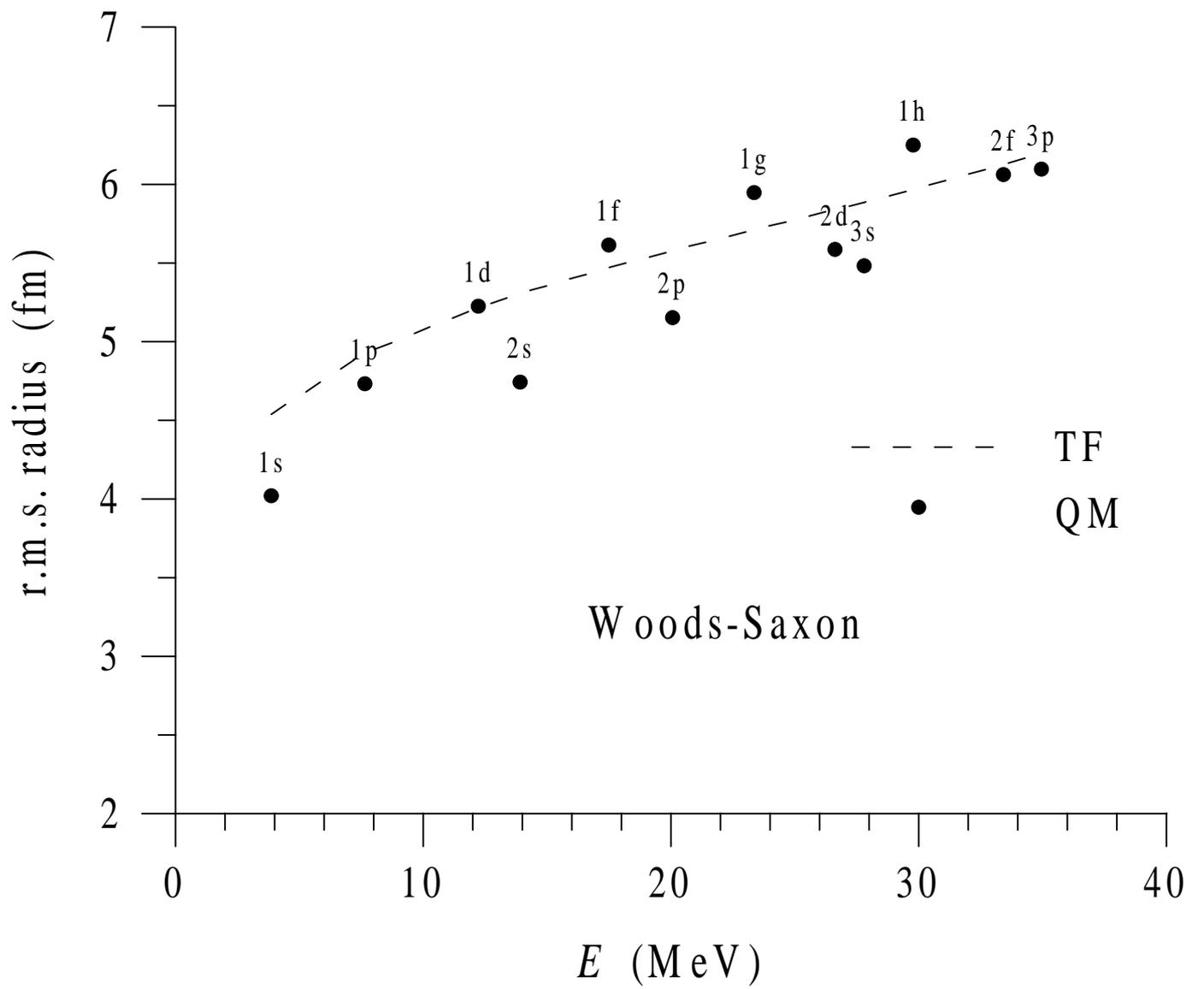
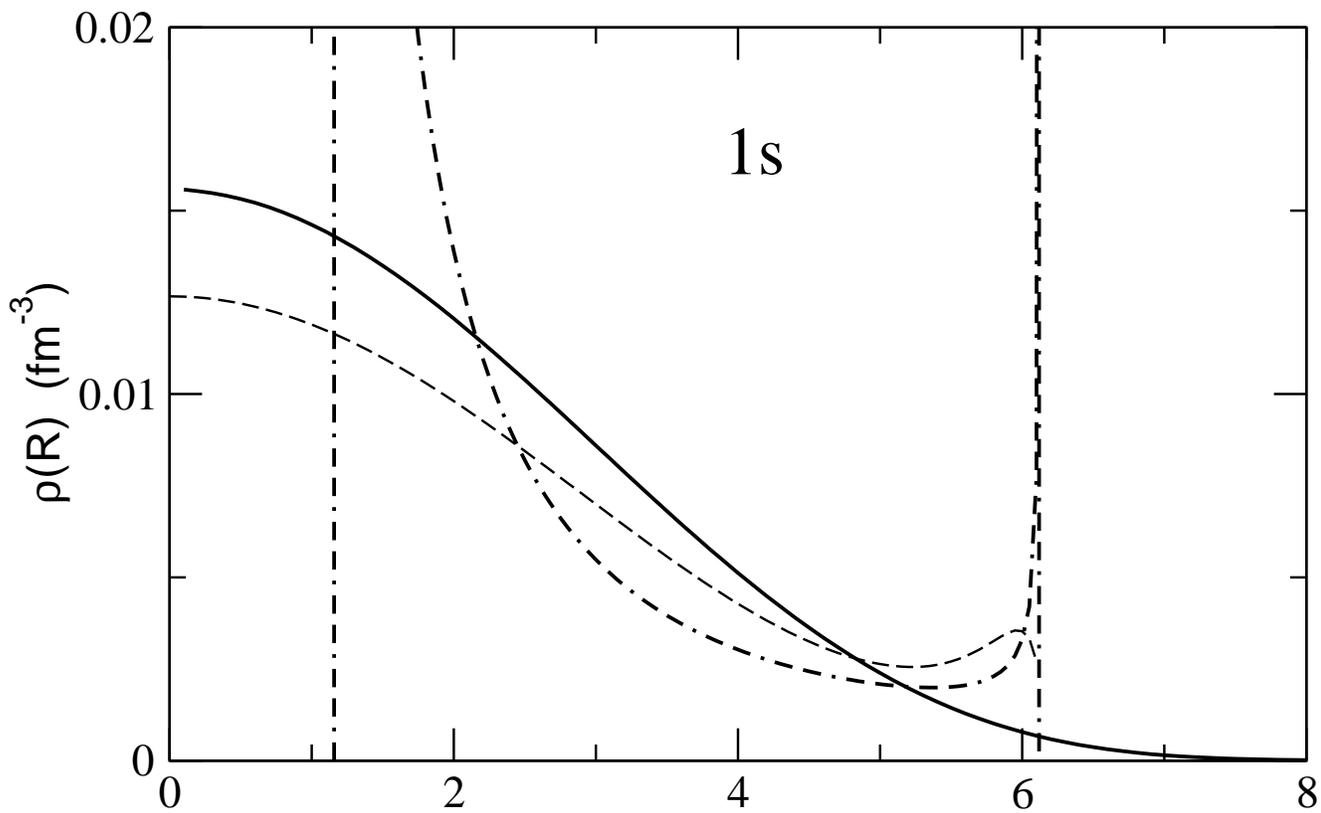
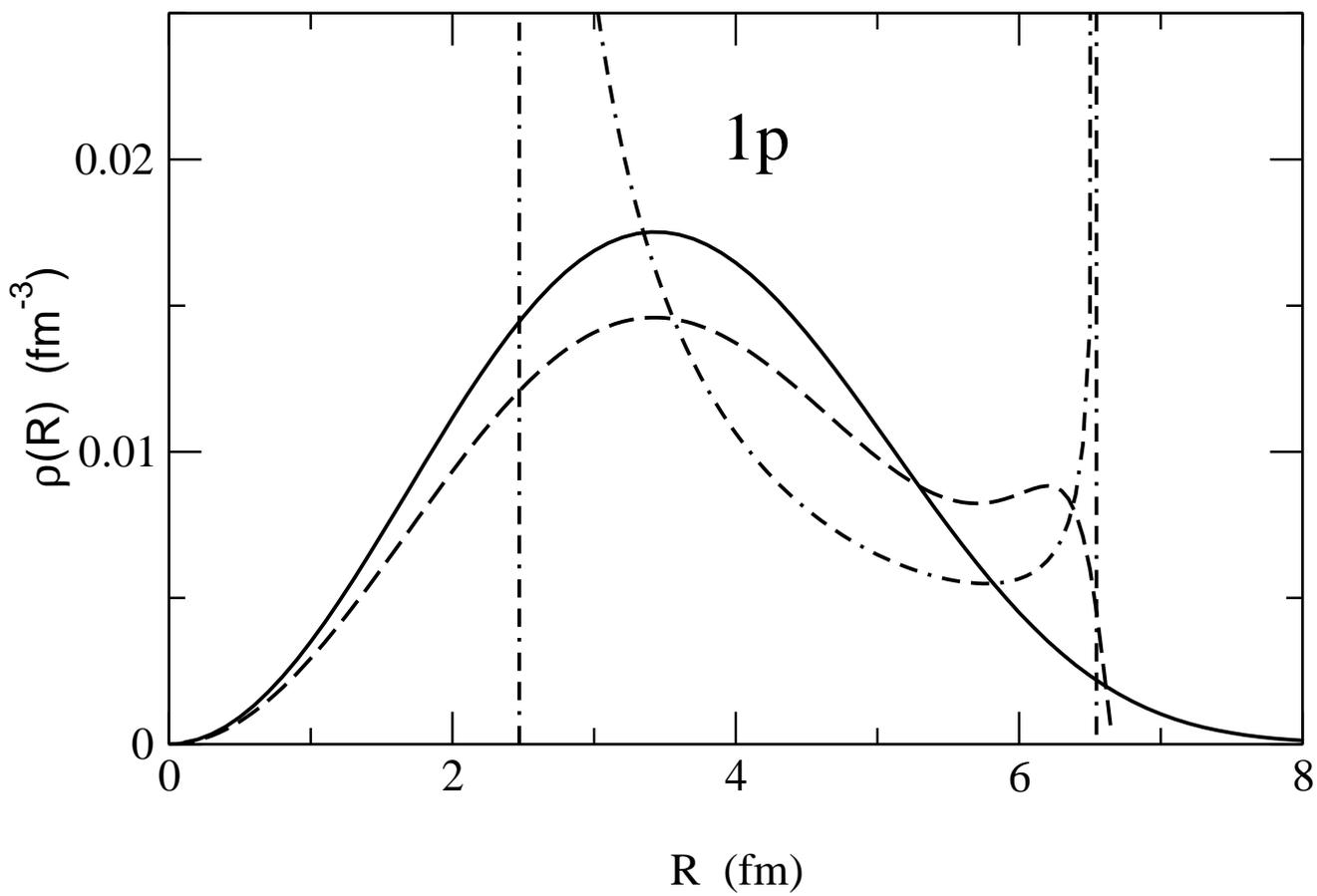
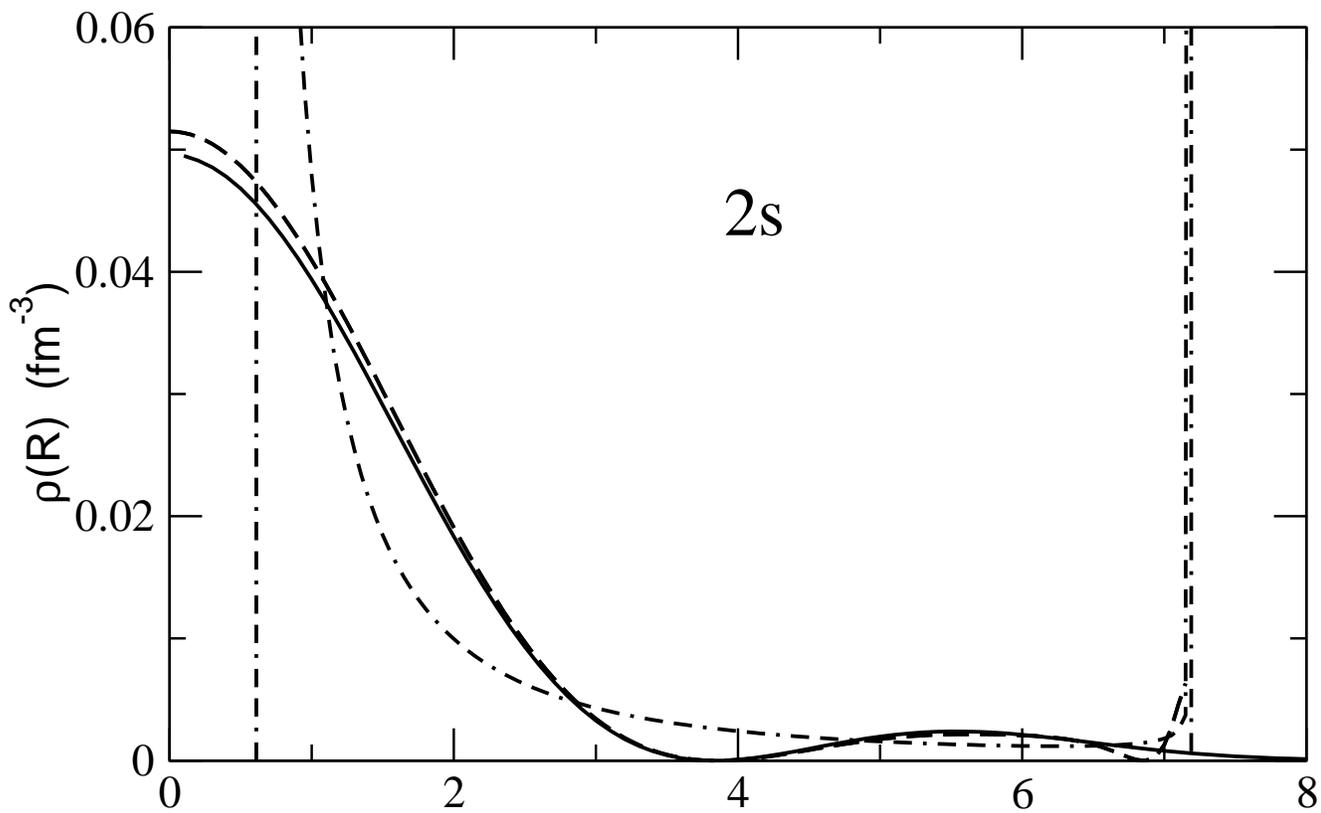
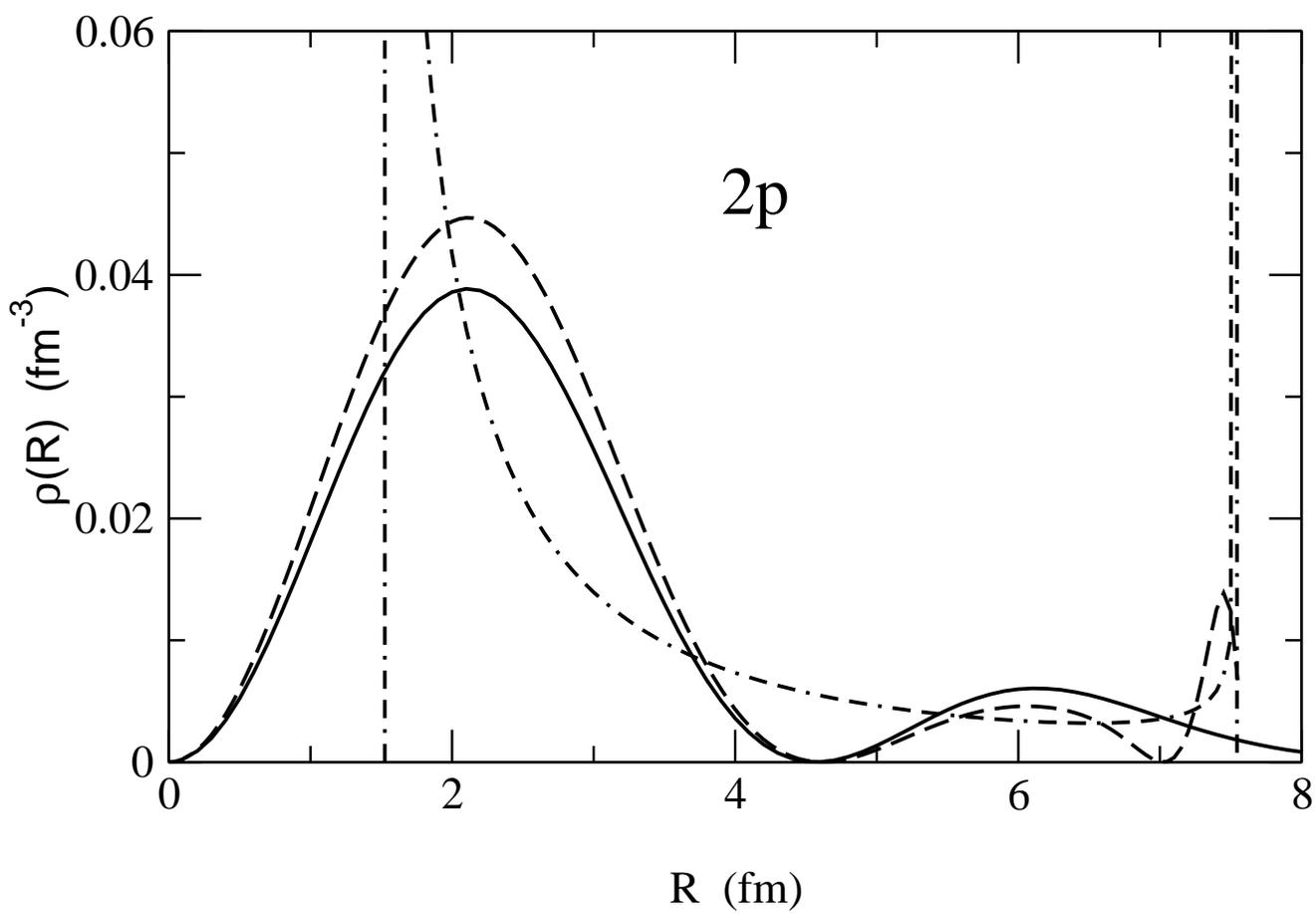
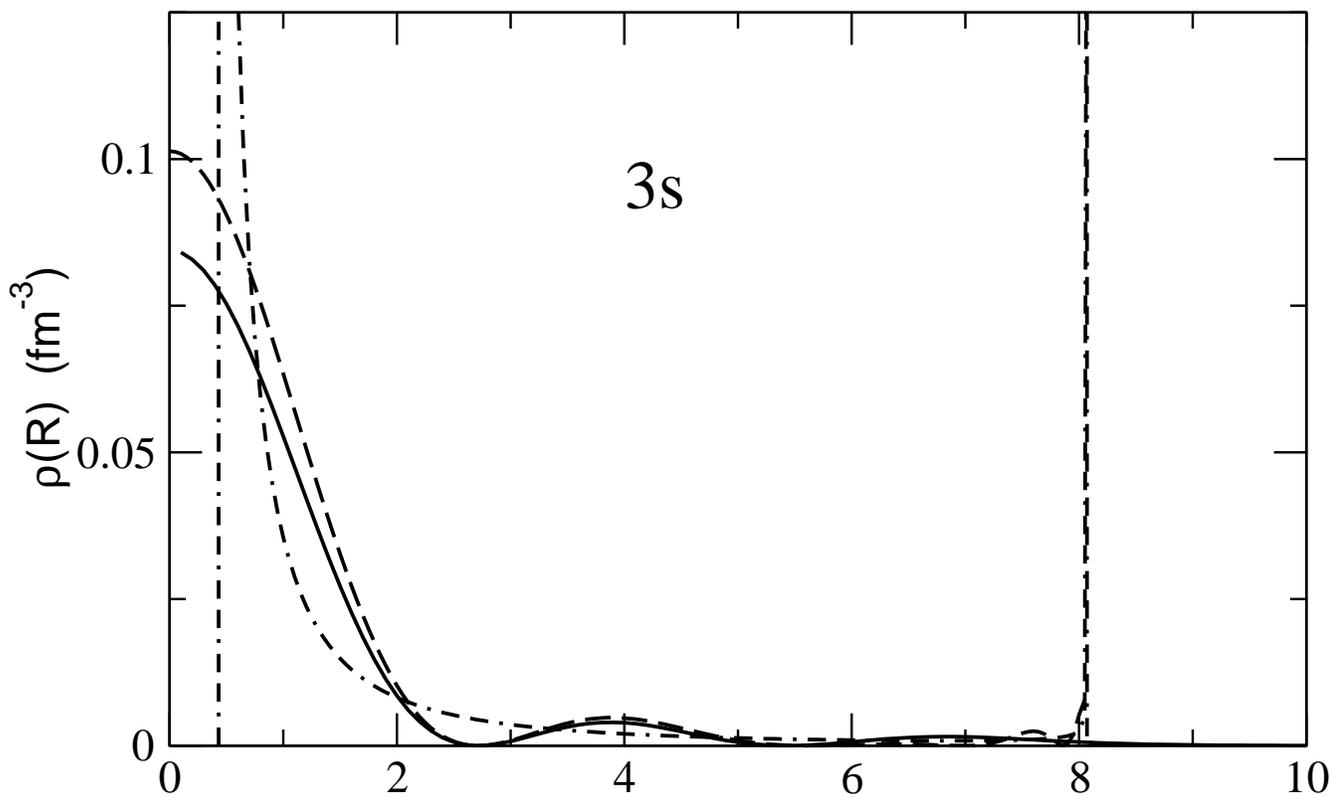
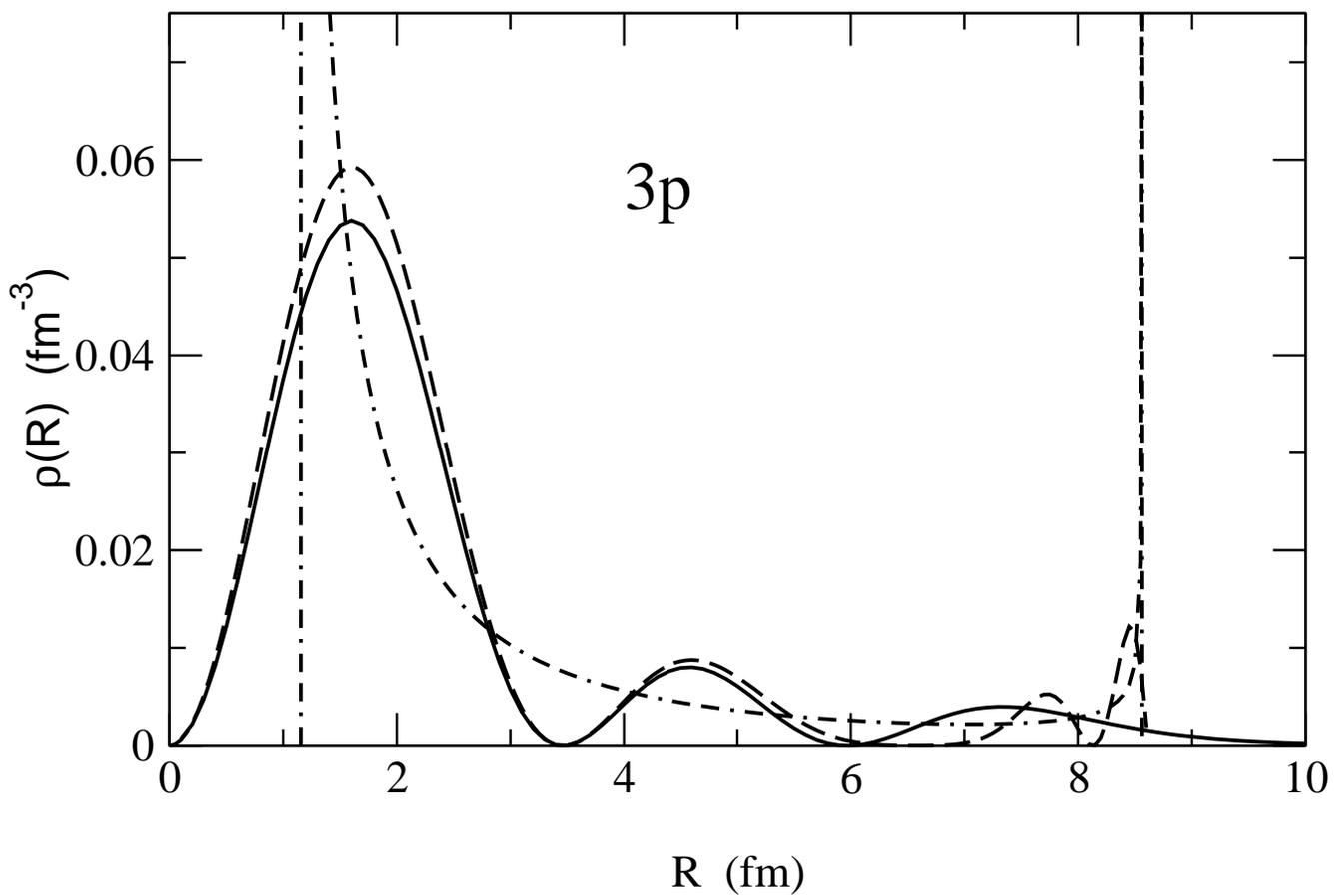
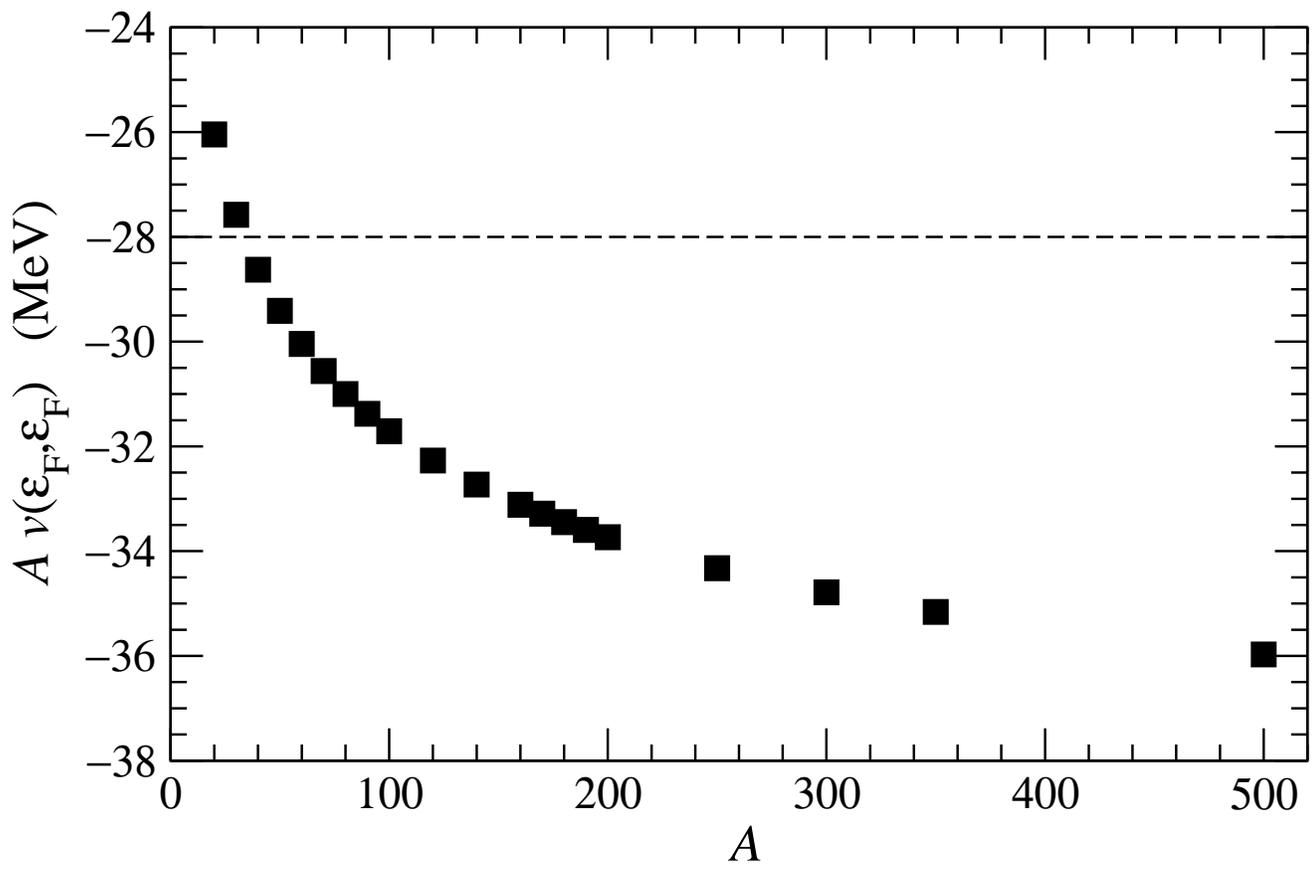
\end{document}